S. V. Charapitsa, PhD (Physics and Mathematics),
head of the Laboratory of analytical research (BSU);
N. I. Zayats, PhD (Engineering), assistant Professor (BSTU);
Y. V. Zadreyko, Vice Head of the Metrology Department of
the State Committee for Standardization of the Republic of Belarus (Gosstandart) ;
S. N. Sytova, PhD (Physics and Mathematics), scientific secretary (BSU)

# ESTIMATION OF THE ACCURACY OF METHOD FOR QUANTITATIVE DETERMINATION OF VOLATILE COMPOUNDS IN ALCOHOL PRODUCTS

Results of the estimation of the precision for determination volatile compounds in alcohol-containing products by gas chromatography: acetaldehyde, methyl acetate, ethyl acetate, methanol, isopropyl alcohol, propyl alcohol, isobutyl alcohol, butyl alcohol, isoamyl alcohol are presented. To determine the accuracy, measurements were planned in accordance with ISO 5725 and held at the gas chromatograph Chromatec-Crystal 5000. Standard deviation of repeatability, intermediate precision and their limits are derived from obtained experimental data. The uncertainty of the measurements was calculated on the base of an "empirical" method. The obtained values of accuracy indicate that the developed method allows measurement uncertainty extended from 2 to 20% depending on the analyzed compound and measured concentration.

**Introduction.** In accordance with the Law "On uniformity of measurements" the measurements should be carried out according to the procedures of measurements (PM) which are certified in a proper manner. The requirements to PM are established by GOST 8.010 and validation is accomplished according to the requirements 8.006 of Technical Code of Practice. The method of measurement is a set of operations and rules, which provides the results with known accuracy that is to be a basic requirement to guarantee uniform measurements. Regarding this, the method of measurements should contain the accuracy values, which can be represented by the correctness and/or precision, assigned characteristics of measurement uncertainty. The accuracy is characterized by bias (deviation from reference value); precision is determined by repeatability (parallel results proximity), intermediate precision (which is determined by proximity of the results obtained in the same laboratory, but in different conditions), and reproducibility (the proximity of the results obtained in different laboratories) [1].

To estimate the accuracy of the results of the measurements, the term uncertainty has been increasingly used; it is the main and globally recognized parameter, characterizing the accuracy of the measurements. The expression of uncertainty in accordance with the specified procedures and guidelines is to be a mandatory condition of the measurements results recognition by international organizations, as well as the requirements to be implemented according to ISO/IEC 17025.

Uncertainty is a parameter associated with the results of measurements characterizing the values range, which could be reasonably attributed to the measured parameter [2]. Uncertainty can be expressed as average quadratic deviation (standard uncertainty) or interval (expanded uncertainty), and calculated according to the method A (on the basis of some experimental data) or according to the method B (on the basis of additional information).

**Main part.** The purpose of this paper is determination of the accuracy of the method for quantitative determination of volatile compounds in alcohol-containing products.

The method establishes a gas-chromatographic method for the determination of the following volatile compounds: acetaldehyde (ethanal), methyl acetate, ethyl acetate, methanol, isopropyl alcohol (2-propanol), propyl alcohol (1-propanol), isobutyl alcohol (2-methyl-1-propanol), butyl alcohol (1-butanol), isoamyl alcohol (3-methyl-1-butanol) [3, 4].

The range of measured mass concentration of methanol is from 13 to 20,000 mg per 1 litre of anhydrous ethyl alcohol (AA); for 2-propanol: from 2 to 2,000 mg; and for all other defined volatile compounds: from 1 to 2,000 mg per 1 litre of AA.

The originality of the method is that the internal standard for the analysis of alcohol-containing products is ethanol, which is contained in the tested products and there is no need to add ethanol to the sample. The results of the analysis are expressed in mg per liter of AA.

Calibration of the chromatograph is to establish the relative response factors (RRF) of the detector to each of the analyzed compounds regarding to the ethanol. The numerical values of the RRF are obtained from the chromatographic data of standard samples with known concentrations of ethanol and analyzed compounds.

Series of experiments have been planned in accordance with the requirements of ISO 5725 (2–4) and carried out to evaluate the metrological characteristics of the proposed method. All the experiments were performed in the Laboratory of of analytical research of Research Institute for Nuclear Problems of Belarusian State University. Analysis of samples was performed on a gas chromatograph Chromatec-Crystal 5000 equipped with a PID.

Standard solutions for calibration of the chromatograph and experimental samples to study the accuracy were prepared by adding separate standard compounds (producer Sigma-Aldrich, Fluka, Germany) in aqueous ethanol mixture (96:4 %). Experimental samples with known concentrations of compounds are necessary for determination of correctness. They were also used to measure the repeatability and intermediate precision. As all experiments were carried out in the same laboratory, the reproducibility of the method was not estimated.

The eight standard solutions S1-S8 were prepared by the gravimetric method. Their mass concentrations of methanol were the following: 13; 23; 53; 63; 103; 1,005; 5,013 and 20,000 mg/l (AA); 2-propanol: 2; 4; 7; 8; 11; 100; 500; 2,000 mg/l (AA) and all other defined compounds: 1; 2; 5; 6; 10; 100; 500 and 2000 mg/l (AA). Concentrations were chosen to overlap the entire range of determining compounds according to PM.

For each sample (level, the number of levels $j = 1, \ldots, 8$, $Y_{ij}$) there were performed 15 series of measurements under intermediate precision conditions (different operators, at different times, $i = 1\ldots15$); 2 results of single measurement ( parallel measurements, $k = 1, 2, Y_{ij1}, Y_{ij2}$).

The arithmetic average ($\overline{Y_{ij}}$) of two single measurements was taken as a result. The results were obtained on a single calibration curve for each compound.

To check the statistical spikes among the results of measurements in the conditions of repeatability, the Cochrane criterion was used obtained under conditions of intermediate precision (Grabbs criterion) [5].

According to the obtained results under the formulas presented in the standard [5], the standard repeatability deviation $S_{r,j}$ was calculated. It takes into account the effect of random factors when performing parallel measurements. As an experiment for the evaluation of intermediate precision was combined with the experiment for evaluation repeatability measurement; and the measurement results in terms of intermediate precision (time (T), operator (O)) were the arithmetic mean of the two parallel results when calculating the standard deviation of the intermediate precision at each level the average results were taken into account as $S_{r,j}$:

$$S_{Ij(TO)} = \sqrt{S_{Lj}^2 + S_{rj}^2}, \qquad (1)$$

where $S^2_{rj}$ is the dispersion of repeatability; $S^2_{Lj}$ is the inter-series dispersion calculated by the formula

$$S^2_{Lj} = \frac{1}{p-1}\sum_{i=1}^{p}(\overline{Y_{ij}} - \overline{\overline{Y_j}})^2 - \frac{S^2_{rj}}{2}, \qquad (2)$$

where $\overline{Y_{i,j}}$ is the arithmetic mean of the two parallel results, $\overline{\overline{Y_j}}$ is the average arithmetic mean of the fifteen series.

Laboratory bias, which is an estimate of the accuracy was calculated by the following formula

$$\hat{\Delta} = \overline{\overline{Y_j}} - \mu, \qquad (3)$$

where μ is an accepted reference value for each individual level.

Analysis of the significance of laboratory bias showed that for most levels it was not significant, indicating that there was no system error during the measurements.

To establish accuracy in the whole range of measured concentrations of compounds on the obtained values of the accuracy rate of eight levels, an attempt was made to establish a functional relationship between the accuracy rate and the measured concentrations. However, the results showed that this correlation dependence with a high coefficient of correlation is absent.

Therefore, the entire concentration range was divided into two sub-ranges within which the accuracy can be considered the same. Fisher's exact test was used to delimit sub-ranges. The maximum value of the relative standard deviation of repeatability and intermediate precision in each sub-range were taken as the relative standard deviation for intermediate precision.

The repeatability and intermediate precision limits were established according to the formulas $r = 2,8 \cdot S_r$ and $r = 2,8 \cdot S_{i(TO)}$. These factors are necessary for the implementation of periodic internal control of accuracy when performing measurements according to the PM.

Standard deviations of repeatability and intermediate precision, as well as their limits (percentage) are given in Table 1.

To estimate the uncertainty of measurements the empirical approach was used, as it allows using already selected PM accuracy (correctness and precision) and to estimate the uncertainty of the method in general [6]. In this case, the standard uncertainty of measurements of the determined compound concentration $u$ is calculated according to the formula

$$u = \sqrt{S_{I(TO)}^2 + b^2}, \qquad (4)$$

where $S_{I(TO)}$ is the standard deviation, characterizing intermediate precision measurements; $b$ is the estimation for the bias.

To estimate the uncertainty of measurements of analyzed compound concentrations, the standard deviation of precision $S_{I(TO)}$ was used as precision factor, because it takes into account more factors affecting the precision with respect to standard deviation of repeatability.

Table 1

**PM Precision Factors**

| Investigated Compounds | Range of Measured Mass Concentrations, mg/l | Standard Deviation of Repeatability, $S_r$, rel. % | Repeatability Limit $r$, rel.% | Standard Deviation of Intermediate Precision, $S_{I(TO)}$, rel. % | Intermediate Precision Limit, $R$, rel. % |
|---|---|---|---|---|---|
| 2-Propanol | From 2 to 10 inc. | 2,3 | 6,4 | 3,0 | 8,4 |
| | From 10 to 2,000 | 0,6 | 1,7 | 0,9 | 2,5 |
| 1-Propanol | From 1 to 10 incl. | 3,8 | 10,6 | 6,0 | 16,8 |
| | From 10 to 2,000 | 1,2 | 3,4 | 1,5 | 4,2 |
| 1-Butanol | From 1 to 10 incl. | 4,4 | 12,3 | 6,3 | 17,6 |
| | From 10 to 2000 | 0,2 | 0,6 | 0,4 | 1,1 |
| Isobutyl Alcohol | From 1 to 10 incl. | 4,0 | 11,2 | 4,5 | 12,6 |
| | From 10 to 2,000 | 0,2 | 0,6 | 0,3 | 0,8 |
| Isoamyl Alcohol | From 1 to 10 incl. | 3,8 | 10,6 | 6,0 | 16,8 |
| | From 10 to 2,000 | 1,2 | 3,4 | 1,3 | 3,6 |
| Methyl Acetate | From 1 to 10 incl. | 3,7 | 10,3 | 3,9 | 10,9 |
| | From 10 to 2,000 | 0,3 | 0,9 | 2,4 | 6,8 |
| Ethyl Acetate | From 1 to 10 incl. | 3,6 | 10,1 | 4,7 | 13,0 |
| | From 10 to 2,000 | 1,3 | 3,6 | 2,2 | 6,2 |
| Acetic Aldehyde | From 1 to 10 incl. | 3,6 | 10,1 | 5,6 | 15,7 |
| | From 10 to 2,000 | 0,7 | 2,1 | 1,5 | 4,2 |
| Methanol | From 13 to 100 incl. | 1,1 | 3,1 | 1,5 | 4,2 |
| | From 100 to 20, 000 | 0,1 | 0,3 | 0,2 | 0,6 |

To estimate the uncertainty of measurements of the analyzed compounds concentrations the standard intermediate precision deviation $S_{I(TO)}$ was used as precision characteristics, because it takes into account more factors affecting the precision compared with standard deviation of repeatability.

The contribution of bias in uncertainty was calculated from the average deviation $\overline{\Delta}$, uncertainty of the reference value $u_{ref}$, and precision of the average value of repeated measurements made in the study of the bias $S_\Delta$ according to the following formula:

$$b = \sqrt{\overline{\Delta}^2 + u^2_{ref} + S_\Delta^2}, \quad (5)$$

the standard deviation in the estimated bias $S_\Delta$ was calculated by the formula:

$$S_\Delta = \sqrt{\frac{\sum_{p=1}^{15}\left(\Delta_{ij} - \overline{\Delta}\right)^2}{p(p-1)}}, \quad (6)$$

where $\Delta_{ij}$ is the bias of results of separate measurements; $\Delta$ is the average arithmetic bias.

To estimate the uncertainty of concentration of the analyzed compound in the prepared solution ($u_{ref}$), the modeling method was used in accordance with the recommendations of the Manual EUROCHEM/SETAC "Quantitative description of uncertainty in analytical measurements" [7]. The method is based on the model determining the measured value (concentration) being affected by other values and determining the affect of each of them in the uncertainty of the measured value.

The measurement model is the functional dependence, which is used to calculate the concentration of the i-th volatile compound in the prepared standard solution.

For example, calculation of mass concentration (mg per 1 litre of anhydrous alcohol) of the i-th volatile compound in the experimental sample $S1$ was carried out according to the following formula:

$$C^i(S1) = \frac{C^i m_{S1}^i + C^i(Et)m_{S1}^{Et}}{(C^{Et}(Et)m_{S1}^{Et} + \sum_{j=1}^{9} C^{Et}(j)m_{S1}^j)/\rho_{Et}}, \quad (7)$$

where $C^i$ is the mass concentration (milligram per 1 mg of solution) of the basic i-th compound in the initial solution of the i-th defined volatile compound, %; $C^i(Et)$ is the mass concentration (milligram per 1 mg of solution) of the i-th compound of the initial ethanol, %; $C^{Et}(Et)$ is the mass concentration (milligrams per 1 mg of solution) of ethanol in the initial ethanol, %; $C^{Et}(j)$ is the mass concentration (milligrams per 1 mg of solution) of ethanol in the initial solutions of the added j-x compounds, %; $m_{S1}^i$ is the mass of the added i-th analyzed volatile compound, mg; $m_{S1}^{Et}$ is mass of the added initial ethanol, mg; $\rho_{Et}$ is the density of anhydrous ethanol, mg/l, under normal conditions; $\rho_{Et} = 789\ 300$ mg/l.

The standard uncertainties of all the values included in the formula (7), were calculated using the uniform distribution law:

$$u(x_i) = \frac{a}{\sqrt{3}}, \quad (8)$$

where $u(x_i)$ is the standard uncertainty of the included values; $a$ is the half interval of measurement uncertainty.

Standard measurement uncertainty was determined by summing the standard uncertainty of the included values (the square root of the sum of squares), taking into account their weight factors (sensitivity coefficients). Weight factors were calculated as partial derivatives of the function with respect to the input value, for example:

$$\frac{\partial C^i(S1)}{\partial m^i_{S1}}.$$

Thus it was obtained the formula for calculating the standard uncertainty of the mass concentration of the analyzed volatile compounds in the solution $S1$:

$$u(C^i(S1)) = \left[\left(\frac{\rho_{Et}C^i - C^i(S1)C_i^{Et}}{Z(S1)}u(m^i_{S1})\right)^2 + \right.$$

$$+\left(\frac{\rho_{Et}m^i_{S1}}{Z(S1)}u(C^i)\right)^2 +$$

$$+\left(\frac{\rho_{Et}C^i(Et) - C^i(S1)C^{Et}}{Z(S1)}u(m^{Et}_{S1})\right)^2 +$$

$$+\left(\frac{\rho_{Et}m^{Et}_{S1}}{Z(S1)}u(C^i(Et))\right)^2 +$$

$$+\left(\frac{C^i(S1)m^{Et}_{S1}}{Z(S1)}u(C^{Et})\right)^2 +$$

$$+\sum_{\substack{j=1\\j\neq i}}^{9}\left(\frac{C^i(S1)C_j^{Et}}{Z(S1)}u(m^j_{S1})\right)^2 +$$

$$+\sum_{j=1}^{9}\left(\frac{C^i(S1)m^j_{S1}}{Z(S1)}u(C_j^{Et})\right)^2\right]^{1/2};$$

$$Z(S1) = C^{Et}(Et)m^{Et}_{S1} + \sum_{j=1}^{9} C_j^{Et} m^j_{S1}$$

(9)

where $u(m^i_{S1})$ is the mass uncertainty of the added i-th analyzed volatile compound, mg; $u(C^i)$ is the mass concentration uncertainty (milligram per 1 mg of solution) of the basic i-th compound in the initial solution of the i-th defined volatile compound, %, it can be calculated by the following formula:

$$u(C^i) = \left(\sum_{j=1(j\neq i)}^{9} u^2(C^j(i))\right)^{1/2}, \quad (10)$$

where $u(C^j(i))$ is the standard uncertainty of the mass concentration (milligram per 1 mg of solution) of the j-th compound in the i-th initial compound, %; $u(m^{Et}_{S1})$ is the standard mass uncertainty of added initial ethanol, mg; $u(C^{Et}(Et))$ is standard uncertainty of mass concentration (milligram per 1 mg of solution) of the i-th compound in the initial ethanol, %; $u(C_j^{Et})$ is the standard uncertainty of the mass concentration (milligram per 1 mg of solution) of ethanol in the initial solutions of j-th added compounds, %.

All of the above standard uncertainties were calculated according to the formula (8).

Table 2 shows the results of calculations of the relative standard uncertainty and extended uncertainty, calculated at confidence coefficient 0.95, and the coverage ratio 2.

Table 2

**Standard and Expanded Uncertainties of Measurements**

| Analyzed Compounds | Range of Measured Mass Concentrations, mg/l | Relative Standard Uncertainty $u$, % | Relative Expanded Uncertainty $U$, %; $P = 0,95$; $k = 2$ |
|---|---|---|---|
| 2-Propanol | From 2 to 10 inc. | 10,0 | 20,0 |
|  | From 10 to 2,000 | 4,0 | 8,0 |
| 1-Propanol | From 1 to 10 incl. | 7,0 | 14,0 |
|  | From 10 to 2,000 | 4,0 | 8,0 |
| 1-Butanol | From 1 to 10 incl. | 9,0 | 18,0 |
|  | From 10 to 2,000 | 3,0 | 6,0 |
| Isobutyl Alcohol | From 1 to 10 incl. | 6,0 | 12,0 |
|  | From 10 to 2,000 | 2,5 | 5,0 |
| Isoamyl Alcohol | From 1 to 10 incl. | 8,0 | 16,0 |
|  | From 10 to 2,000 | 4,0 | 8,0 |
| Methyl Acetate | From 1 to 10 incl. | 10,0 | 20,0 |
|  | From 10 to 2,000 | 5,0 | 10,0 |
| Ethyl Acetate | From 1 to 10 incl. | 8,0 | 16,0 |
|  | From 10 to 2,000 | 4,0 | 8,0 |
| Acetic Aldehyde | From 1 to 10 incl. | 7,0 | 14,0 |
|  | From 10 to 2,000 | 5,0 | 10,0 |
| Methanol | From 13 to 100 incl. | 10,0 | 20,0 |
|  | From 100 to 20,000 | 2,0 | 4,0 |

The table presents data on the measurements precision showing that the developed technique allows implementing measurements with the expanded uncertainty for different volatile compounds from 2 to 20%.

**Conclusion.** The experimental investigations were planned and carried out in accordance with ISO 5725 (2-4). The results of investigations allowed us to determine the accuracy of the new method of determination of impurities in vodka and ethyl alcohol. In 2013, the certification was completed in the Federal Agency for Technical Regulation and Metrology of the Russian Federation for method of measurement to determine the composition of volatile compounds in alcohol and alcohol-containing products (certificate No. 253.0169/01.00258/2013).

**References**

1. Accuracy (trueness and precision) of measurement methods and results: STB ISO 5725-1-2002. Part 1: General principles and definitions. Minsk: BelGISS, 2003. 25 p.

2. Guide to the Expression of Uncertainty in Measurement. SPb .: D.I. Mendeleev VNIIM, 1999. 126 p.

3. Ethanol as an internal standard for a quantitative analysis of toxic impurities in spirit drinks/ S.V. Charapitsa [et al.] // Reports of the National Academy of Sciences. 2012. T. 56, № 1. pp 65-70.

4. Quantitative determination of trace impurities in alcohol production using ethanol as an internal standard / S.V. Charapitsa [et al.] // Food Industry: Science and Technology. 2012. № 2 (16). S. 86-94.

5. Accuracy (trueness and precision) of methods and results of measurements: STB ISO 5725-2-2002. Part 2: Basic method for the determination of repeatability and reproducibility of a standard measurement method. Minsk: BelGISS, 2003. 53 p.

6. Kachur S.A. Alternative approaches to the estimation of measurement uncertainty (based on the technical report EUROLAB, №1 / 2007): method. Manual. Minsk: BelGIM, 2011. 65 p.

7. Quantitative Description of Uncertainty in Analytical Measurements. Guide EUROCHEM/ SETAC. SPb.: D.I. Mendeleev VNIIM, 2002. 149 p.